\documentclass[prl,aps,twocolumn]{revtex4}
\usepackage[latin1]{inputenc} \usepackage{amsmath}
\usepackage{amsfonts} \usepackage{amssymb} \usepackage{graphicx}
\usepackage{color} \usepackage{hyperref} \usepackage{xspace}
\usepackage{dcolumn}
\usepackage{booktabs}  

\newcommand{\dd}{{\rm d}} 

\newcommand{\UNIT}[1]{\ensuremath{\,{\rm #1}}\xspace}

 \newcommand{\GeV}{\UNIT{GeV}}

 \newcommand{\fm}{\UNIT{fm}}
 
\newcommand{\fmc}{\ensuremath{\,{\rm fm}/c}\xspace}

\definecolor{magenta}{cmyk}{0,1,0,0}

\begin{document}

\title{Thermalization of Hadrons via Hagedorn States }

\author{M.~Beitel} \author{K.~Gallmeister} \author{C.~Greiner}
\affiliation{Institut f\"ur Theoretische Physik, Goethe-Universit\"at
  Frankfurt am Main, Max-von-Laue-Str.~1, 60438 Frankfurt am Main,
  Germany}

\begin{abstract}
  Hagedorn states are characterized by being very massive hadron-like
  resonances and by not being limited to quantum numbers of known
  hadrons. To generate such a zoo of different Hagedorn states, a
  covariantly formulated bootstrap equation is solved by ensuring
  energy conservation and conservation of baryon number $B$,
  strangeness $S$ and electric charge $Q$. The numerical solution of
  this equation provides Hagedorn spectra, which enable to obtain the
  decay width for Hagedorn states needed in cascading decay
  simulations. A single (heavy) Hagedorn state cascades by various
  two-body decay channels subsequently into final stable hadrons.  All
  final hadronic observables like masses, spectral functions and decay
  branching ratios for hadronic feed down are taken from the hadronic
  transport model UrQMD. Strikingly, the final energy spectra of
  resulting hadrons are exponential showing a thermal-like
  distribution with the characteristic Hagedorn temperature.
\end{abstract}

\maketitle

In the 60's of the last century physicists were puzzled by the
diversity of different hadron species growing with beam energy. Before
the emergence of quantum chromodynamics (QCD) as the theory of strong
interactions many ideas came up to explain these findings.
R.~Hagedorn \cite{Hagedorn:1965st} proposed to describe the variety of
particles found by a common mass spectrum, now better known as
Hagedorn spectrum, arising in the framework of a "statistical
bootstrap model". In the infinite mass limit this spectrum is
exponentially rising where the slope is determined by the Hagedorn
'temperature' $T_H$.  Above this temperature the partition function of
a strongly interacting hadronic system with such an exponential growth
of states diverges and a new state of matter, the 'Quark Gluon Plasma'
(QGP), is assumed to be realized. One of the most challenging problems
is to understand how this phase transition exactly occurs and which
new properties this new state of matter has.  One possible tool to
investigate microscopically a phase transition from hadronic to
partonic phase is the application and generation of Hagedorn states
being created in multiparticle collisions
~\cite{Greiner:2000tu,Greiner:2004vm,
  NoronhaHostler:2007jf,NoronhaHostler:2010dc,NoronhaHostler:2009cf}.
These resonances belong to the continuous part of the Hagedorn
spectrum and are allowed to have any mass larger than that of the
heaviest known hadron and also any quantum numbers as long as they are
compatible with their mass. Such Hagedorn states can alter the
occurrence of various phases from hadronic to deconfined partonic
matter
\cite{Moretto:2005iz,Zakout:2006zj,Zakout:2007nb,Ferroni:2008ej,Bugaev:2008iu,Ivanytskyi:2012yx}.

The abundant appearance of Hagedorn states in the vicinity of $T_H$
helps to explain how chemical equilibrium of hadrons is achieved on
timescales significantly smaller than the typical lifetime of a
fireball $\left(t\approx 10\fmc\right)$. In
Refs.~\cite{NoronhaHostler:2007jf,NoronhaHostler:2010dc,NoronhaHostler:2009cf},
the authors examined chemical equilibration times of (multi-) strange
(anti-) baryons at Relativistic Heavy Ion Collider (RHIC) energies by
solving a set of coupled rate equations. It was assumed, that most
abundant mesons (pions, kaons) 'cluster' to Hagedorn states which in
turn decay into hyperons driving them quickly into equilibrium. For
example, the chemical equilibration times of protons, kaons and
lambdas within this approach are of the order of 1-2\fmc. The
inclusion of Hagedorn states in a hadron resonance gas model provides
a lowering of the speed of sound, $c_s$, at the phase transition and
being in good agreement with lattice calculations
\cite{NoronhaHostler:2008ju,Majumder:2010ik,NoronhaHostler:2012ug,Jakovac:2013iua}.
In addition, by comparing calculations with inclusion of Hagedorn
states to calculations without them, a significant lowering of the
shear viscosity to entropy ratio, $\eta/s$, is observed
\cite{NoronhaHostler:2008ju,Gorenstein:2007mw,Itakura:2008qv,NoronhaHostler:2012ug}.
The inclusion of Hagedorn states creates a minor dependence of the
thermal fit parameters of particle ratios on the Hagedorn temperature,
$T_H$, which is assumed to be equal to $T_C$
\cite{NoronhaHostler:2009tz}.

In order to describe the hadronization of jets in
e$^+$e$^-$--annihilation events, different scenarios were developed
during the 70's and 80's of the last century: While the first one
assumes independent parton fragmentation \cite{Field:1977fa}, the
fundamental objects of the second approach are color strings
\cite{Andersson:1983ia}.  Finally, the basic assumption of the latter
is that partons tend to cluster in color singlet states from the very
beginning of the generated event.  These cluster then decay to smaller
ones, until some cut-off scale is reached and hadrons are formed
\cite{Webber:1983if,Marchesini:1991ch}.
An explicit application of the "statistical bootstrap model" has been
used to calculate several properties of particles stemming from decays
of hadronic fireballs being created in relativistic heavy ion
collisions \cite{Hagedorn:1980kb}.  In the framework of RQMD
multi-particle collisions and their decays were introduced by the so
called particle clusters a particle system can separate into provided
the existence of separable interactions in the relativistic particle
dynamcis exists.  This clustering is for example fulfilled for colored
quarks and gluons \cite{Sorge:1989dy}.  Another statistical approach
within the microcanonical ensemble addressed the hadronization of
quark matter droplets \cite{Werner:1995mx}.  A further statistical
treatment of Hagedorn states was performed in Ref.~\cite{Pal:2005rb}
by forcing detailed balance between creation and decay of Hagedorn
states with a simplistic description of the spectrum in the low mass
region. The authors made then the extreme assumption of one single
heavy Hagedorn state subsequently decaying into stable hadrons giving
rise to measured particle multiplicities at RHIC and Super Proton
Synchrotron (SPS) energies.
The several terms like 'quark matter droplets', 'clusters' or
'fireballs' may all be identified with possible Hagedorn states.

The present work formulates the whole zoo of Hagedorn states and their
decay properties, as created in binary collisions within the
microscopic hadronic transport simulation program UrQMD
\cite{Bass:1998ca}.  Multiplicities (and their ratios) of stable
hadrons stemming from cascading decay simulations of massive Hagedorn
states are calculated.  Additionally energy distribution of the decay
products are examined and it is shown that all hadrons stemming from
that cascade follow the Boltzmann distribution resulting in a
thermalized hadron resonance gas.
Contrary to the well-known non-covariant bootstrap equation
\cite{Frautschi:1971ij}, here a covariantly formulated \cite{BycKaj73}
bootstrap equation
\begin{align}\label{eq:tautotbsq}
  &\tau_{\small{\vec{C}}}\left(m\right)=\frac{R^3}{3\pi
    m}\sum\limits_{\small{\vec{C}_1,\vec{C}_2}}\iint
  \dd m_1\dd m_2\,\tau_{\small{\vec{C}_1}}(m_1)\,m_1\\
  &\times\tau_{\small{\vec{C}_2}}(m_2)\,m_2\
  p_{cm}\left(m,m_1,m_2\right)\delta^{\left(3\right)}\left(\vec{C}-\vec{C}_1-\vec{C}_2\right)\
  ,\nonumber
\end{align}
is used, which ensures strict energy and quantum number conservation,
$\vec{C}=\left(B,S,Q\right)$. ($R$ stands for the radius of the
Hagedorn state's volume as discussed below.)  The main idea behind any
such bootstrap equation is the "statistical bootstrap model" which
postulates that fireballs consist of fireballs which in turn consist
of fireballs etc., resulting in a common spectrum with the remarkable
feature that it grows approximately exponentially in the infinite mass
limit.  The present approach is restricted to two constituents only
making up a Hagedorn state because the focus is put on
$2\leftrightarrow1$ processes only.  With this the principle of
detailed balance can be applied, which can be implemented into a
standard hadronic transport framework.  This restriction is backed by
the Hagedorn state decay probability into $n$ particles,
$P\left(n\right)=\left(\ln 2\right)^{n-1}/\left(n-1\right)!$, yielding
a probability for the decay into two particles of $69\%$, into three
particles of $24\%$ etc.~\cite{Frautschi:1971ij}.
The bootstrap equation (\ref{eq:tautotbsq}) in general is a highly
non-linear integral equation of Volterra type which can be solved
analytically for some special cases
\cite{Yellin:1973nj,Hagedorn:1973kf}.
Here the basic input are the spectral functions provided by the
hadronic table of URQMD consisting of 55 different baryons and 32
different mesons \cite{Bass:1998ca}, calling for a numerical solution.

Thus one starts by inserting known hadron spectral functions on the
r.h.s.~of Eq.~(\ref{eq:tautotbsq}) resulting into first Hagedorn states
on the l.h.s.~of this equation.  In the subsequent steps, these
created Hagedorn states serve as constituents in addition to the known
sources.  In each step, quantum number conservation
$\vec{C}=\vec{C}_1+\vec{C}_2$ is assured.  In this fashion one
proceeds by increasing the mass by steps of $\Delta
m=0.01\GeV$. Unfortunately, the computation time increases with mass
according $m^8$, since more and more constituents have to be taken
into account. Thus the applicability of this approach is limited to
the region $m\le8\GeV$.

Given the Hagedorn spectra, $\tau_{\small{\vec{C}}}\left(m\right)$,
one is able to derive a formula for the total decay width of the
Hagedorn states. For this purpose one modifies the general two body
decay formula \cite{PhysRevD.86.010001ab} to take the mass
degeneration into account. In the general formulae for cross section
and decay width, the creation, $\left|\mathcal{M}_c\right|^2$, and the
decay matrix elements, $\left|\mathcal{M}_d\right|^2$, appear which
for Hagedorn states are at first unknown. By demanding the principle
of detailed balance
$\left|\mathcal{M}_c\right|^2$=$\left|\mathcal{M}_d\right|^2$, one
eventually leads to
\begin{align}\label{eq:gamgen}
  &\Gamma_{\vec{C}}\left(m\right)=\frac{\sigma}{2\pi^2\tau_{\vec{C}}\left(m\right)}\sum\limits_{\vec{C}_1,\vec{C}_2}\iint\dd m_1\dd m_2\tau_{\vec{C}_1}\left(m_1\right)\tau_{\vec{C}_2}\left(m_2\right)\nonumber\\
  &\times
  p_{cm}\left(m,m_1,m_2\right)^2\delta^{\left(3\right)}\left(\vec{C}-\vec{C}_1-\vec{C}_2\right)\
  .
\end{align}
Here, by connecting the radius parameter $R$ of (\ref{eq:tautotbsq})
with the cross section $\sigma$ of (\ref{eq:gamgen}) via $\sigma=\pi
R^2$ the size of the Hagedorn state is connected with its production
and decay properties.  This decay width formula of the Hagedorn state
provides one with the various two-body branching ratios needed for
calculation of hadronic multiplicities in cascade simulations.

\begin{figure}
  \centering
  \includegraphics[width=0.45\textwidth,angle=270]{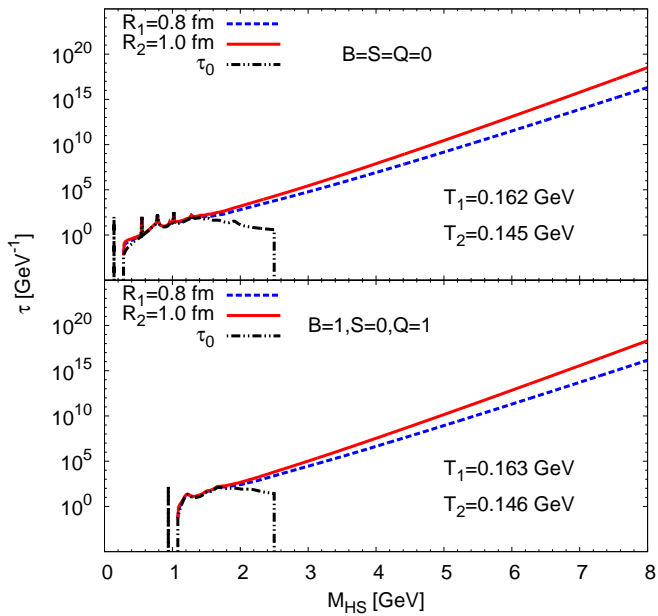}
  \caption{Mesonic $\left(B=S=Q=0\right)$ (up) and baryonic
    $\left(B=1,S=0,Q=1\right)$ (down) Hagedorn spectra for two
    different radii with corresponding (fitted) Hagedorn
    temperatures. The black line represents the sum of spectral
    functions of hadrons with the given quantum numbers.}
  \label{fig:tauprlst}
\end{figure}

The numerical solution of the given bootstrap equation for a mesonic,
non-strange and electrically neutral (B=S=Q=0) Hagedorn spectrum for
two different typical radii ($R_1=0.8\fm, R_2=1.0\fm$) is presented in
Fig.~\ref{fig:tauprlst}.  In the same figure also spectra for baryonic
non-strange and electrically charged states $\left(B=1,S=0,Q=1\right)$
are shown.
All Hagedorn spectra rise exponentially for masses $\geq 1.5\GeV$ with
different slopes for different radii, but for $m<1.5\GeV$ they all
include and thus fit the 'hadronic' part of the spectrum. Here lies
the major advantage of the present approach since ad hoc assumptions
of the kind $\tau\left(m\right)=f\left(m\right)\exp\left(m/T_H\right)$
with most used pre-functions $f\left(m\right)=Am^{-b}$ or
$f\left(m\right)=A\left(m^2+m_r^2\right)^{-b}$ fail to describe the
low mass region of the spectrum. The slopes of the exponential part
depend strongly on the size of the Hagedorn state since in a larger
one more states can be counted than in a smaller one. The slope
parameter is the well known Hagedorn temperature $T_H$ being extracted
with the fit function $\tau_{\rm
  fit}\left(m\right)=Am^{-b}\exp\left(m/T_H\right)$, yielding
$T_H=0.145\GeV$ for $R=1.0\fm$ and $T_H=0.162\GeV$ for $R=0.8\fm$.
Thus smaller Hagedorn states exhibit a
larger Hagedorn temperatures depending on the energy density. The
Hagedorn temperature range is basically the same for mesonic and baryonic
spectra in our model in contrast to \cite{Broniowski:2004yh} where
'mesonic' and 'baryonic' Hagedorn temperatures differ significantly
because they were extracted not from the continuous part of the
Hagedorn spectrum but from its low-mass region.

\begin{figure}
  \centering
  \includegraphics[scale=0.35,angle=270]{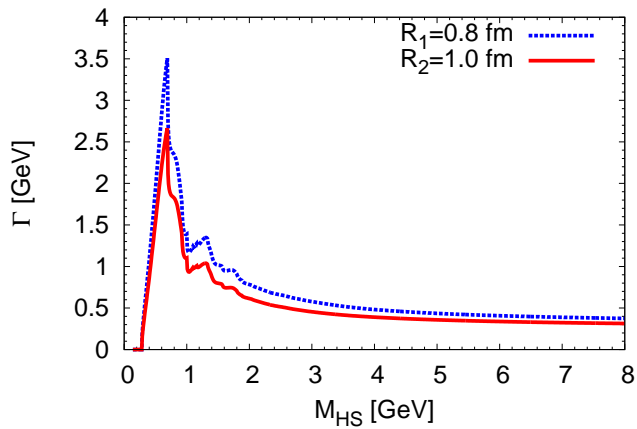}
  \caption{Total decay width of charge neutral Hagedorn state for two
    different radii.}
  \label{fig:gamprl}
\end{figure}

In Fig.~\ref{fig:gamprl} the total decay width of a mesonic,
non-strange and electrically uncharged (B=S=Q=0) Hagedorn state for
same two different radii as before is shown.
The total decay width of a Hagedorn state consists of three different
contributions, where the first one considers only hadrons, 
the second hadrons and Hagedorn
states, and the third one only
Hagedorn states in the outgoing
channel. The peak in the mass range of $M_{HS}=$0-2\GeV comes mainly
from the first contribution, because in this mass range the phase
space for pure hadronic decay is largest \cite{Beitel2014}. 
The height of the peak
depends on the number of hadronic pairs which quantum numbers all sum
up to the quantum number of the Hagedorn state they are building up,
being large for $B=S=Q=0$. Another remarkable feature is that for both
radii the total decay width tends to a constant value depending only
on $R$. This finding is expected by causality where the lifetime of
resonance of dimension $R$ against decay should be roughly the
light-travel time across $R$.

Having the numerous branching ratios at hand, one is able to
calculate hadronic multiplicities stemming from Hagedorn state decays.
Here one starts with some initial heavy Hagedorn state
which decays subsequently down until hadrons are left only. Among
those also non stable resonances might appear which further undergo a
hadronic feed down leaving us with light and stable hadrons with
respect to the strong force like pions, kaons, etc.. All hadronic
properties used here were taken from the transport model UrQMD
\cite{Bass:1998ca}.

\begin{figure}
  \centering
  \includegraphics[width=0.45\textwidth,angle=270]{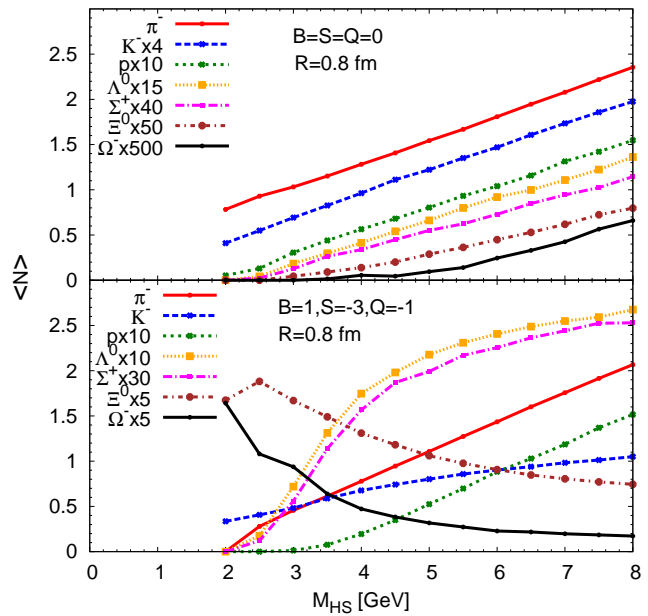}
  \caption{Hadronic multiplicities after a cascade decay of Hagedorn
    state with radius $R=0.8\fm$ and $B=S=Q=0$ (up) and
    $B=1,S=-3,Q=-1$ (down) and the following hadronic feed down.}
  \label{fig:mulprlst}
\end{figure}

Calculated multiplicities for some uncharged
$\left(B=S=Q=0\right)$ initial Hagedorn state are shown in
Fig.~\ref{fig:mulprlst}.
One observes a linear dependence of all multiplicities on the initial
Hagedorn state mass where the magnitude depends on the available phase
space for each hadron. Thus in a decay of a charge neutral Hagedorn
state $\pi^-$ dominate which have to be produced in pairs mostly with
$\pi^+$ since exact charge conservation is enforced. Kaons, especially
$K^-$, are even stronger suppressed not only of their larger mass but
also due to the fact that they have to conserve both electric charge
and strangeness. For the baryons presented the same argumentation
holds since both have to conserve baryon number $B$ and additionally
electric charge $Q$ for proton and strangeness $S$ for $\Lambda$. For
the multistrange hyperons $\Xi^0$ and $\Omega^-$ the production
suppression is even stronger.  This has to be contrasted with the
results for a baryonic, multi-strange and electrically charged
$\left(B=1,S=-3,Q=-1\right)$ $\Omega^-$-like Hagedorn state also shown
in Fig.~\ref{fig:mulprlst}.
Now the choice of Hagedorn state's initial quantum numbers is
reflected in the preference of baryon production although they are
much heavier than the presented mesons. Especially the abundance of
hyperons $\left(\Omega^-,\Xi^0\right)$ compared to the case discussed
before is striking since the easiest way to conserve the initial
quantum numbers is the production of one $\Omega^-\pi^0$- or one
$\Xi^0K^-$ pair where on the other hand the phase space for all other
hadrons with different quantum numbers is suppressed now. Hence exact
conservation of quantum numbers always causes a competition between
hadron's phase space and its quantum numbers.

\begin{figure}
  \centering
  \includegraphics[scale=0.30,angle=270]{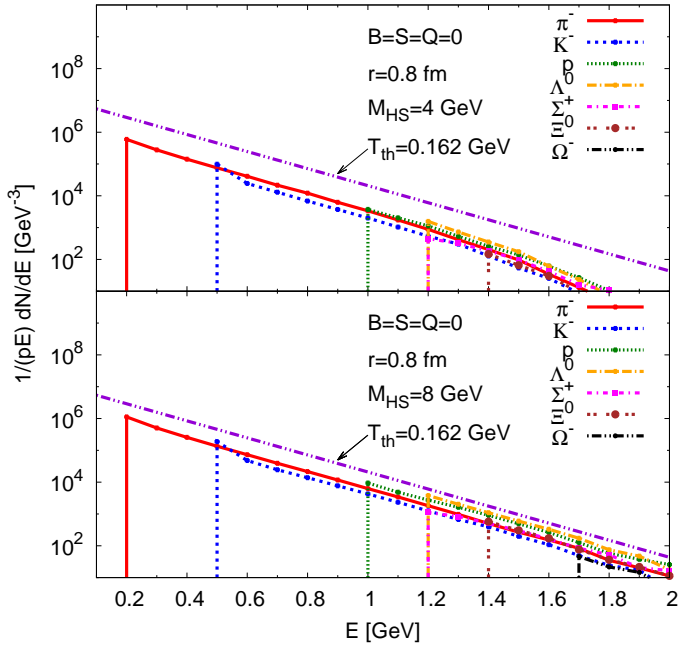}
  \caption{Energy spectra of hadrons stemming from cascade decay of
    charge neutral Hagedorn state with radius $R=0.8\fm$ and initial
    mass $M_{HS}=4\GeV$ and $M_{HS}=8\GeV$.}
  \label{fig:eneprl}
\end{figure}

The energy distribution of
hadrons stemming from Hagedorn state decays in these cascading decay
simulations are some further striking result. 
They are shown in Fig.~\ref{fig:eneprl} for an uncharged $\left(B=S=Q=0\right)$ Hagedorn state with initial
mass $M_{HS}=4\GeV$ and also $M_{HS}=8\GeV$.
\begin{figure}  
  \centering
  \includegraphics[scale=0.35,angle=270]{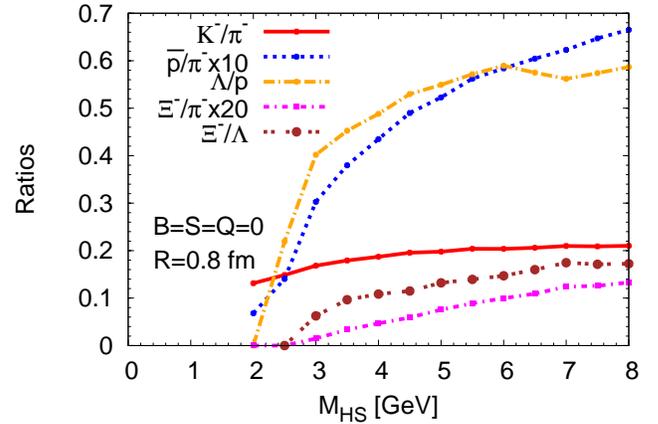}
  \caption{Hadronic ratios stemming from cascade decay of charge
    neutral Hagedorn state with radius $R=0.8\fm$.}
  \label{fig:ratprl}
\end{figure}
The energy
distributions for all species presented follow some exponential
law with the same slope being independent on Hagedorn state's initial
mass. Thus the energies of these final hadrons are distributed akin to
Boltzmann which in turn means that their distribution obey a `thermal`
microcanonical state at a temperature being $T_{th}=0.162\GeV$. This
is remarkable, since this is exactly the Hagedorn temperature
(cf.~Fig.~\ref{fig:tauprlst}). The Hagedorn temperature $T_H$ was
nothing but a slope parameter to fit the exponential part of the
Hagedorn spectrum, where on the other hand $T_{th}$ is a physical of
the created hadron resonance gas. We started with a bootstrap formulae
with no introduction of temperatures at all and obtain a `thermalized´
decay with a temperature being the Hagedorn temperature.

In Fig.~\ref{fig:ratprl} various ratios of most
interesting stale hadrons stemming from a decay of an uncharged
$\left(B=S=Q=0\right)$ Hagedorn state with $R=0.8\fm$ are presented.
Numerical values for the multiplicity ratios for Hagedorn state mases
of 4\GeV and 8\GeV are listed in table \ref{tab:ratios} and compared
to experimental results from ALICE at LHC
\cite{Abelev:2013vea,Abelev:2013xaa,Abelev:2013zaa}. 
\begin{table}[h] 
\begin{ruledtabular} 
\begin{tabular}{cccc} &
 experiment & $4\GeV$ & $8\GeV$\\ 
\hline 
$K^-/\pi^-$ & 0.149 $\pm$ 0.016 & 0.187 & 0.210 \\ 
$\overline p/\pi^-$ & 0.045 $\pm$ 0.005 & 0.043 & 0.066 \\ 
$\Lambda/\pi^-$ & 0.036 $\pm$ 0.005 & 0.021 & 0.038 \\ 
$\Lambda/\overline p$ & 0.778 $\pm$ 0.116 & 0.494 & 0.579 \\ 
$\Xi^-/\pi^-$ & 0.0050 $\pm$ 0.0006 & 0.0023 & 0.0066  \\
$\Omega^-/\pi^-$ & (8.7 $\pm$ 1.7)$\cdot10^{-4}$ & 0.86$\cdot10^{-4}$&
 5.60$\cdot10^{-4}$ 
\end{tabular} 
\end{ruledtabular} 
\caption{Comparison
  of particle multiplicity ratios from theory vs.~experiment
  \cite{Abelev:2013vea,Abelev:2013xaa,Abelev:2013zaa}. Calculated
  values are listed for Hagedorn state masses of 4\GeV and 8\GeV.}
\label{tab:ratios} 
\end{table}

In smaller systems like $e^+$-$e^-$ or $p$-$p$ lighter color
neutral blobs or clusters may be created which solely decay
\cite{Webber:1983if,Marchesini:1991ch}. 
For such small systems one had employed thermal descriptions
incorporating a strangeness suppression factor $\gamma_s$
\cite{Becattini:1995if}.
On the other hand, in relativistic heavy ion collisions larger objets
may be generated which then also interact and are decaying and regenerated.
This may lead to a faster equilibration close to the phase transition
\cite{NoronhaHostler:2007jf,NoronhaHostler:2009cf}. 

Summarizing, such a finding gives new insight into the microscopic and thermal-like
hadronization in ultrarelativistic $e^+$-$e^-$ (see
eg.~\cite{Becattini:1995if}), hadron-hadron-, and also especially in
heavy ion collisions: An implementation of the presented Hagedorn state
decays in addition to their production mechanisms into the transport
approach UrQMD offers a new venue for allowing hadronic multiparticle
collisions in a consistent scheme being important in the vicinity of
the deconfinement transition.  Understanding faster thermalization and
chemical equilibration, but also microscopic transport properties can
be thoroughly investigated in future \cite{Beitel2014}.

This work was supported by the Bundesministerium f\"ur Bildung und
Forschung (BMBF), the HGS-HIRe and the Helmholtz International Center
for FAIR within the framework of the LOEWE program launched by the
State of Hesse. Numerical computations have been performed at the
Center for Scientific Computing (CSC).


\end{document}